\documentstyle[12pt]{article}

\def\Lc{{\cal L}}

\font\tenbb=msym10
\font\sevenbb=msym7
\font\fivebb=msym5
\newfam\bbfam
\textfont\bbfam=\tenbb \scriptfont\bbfam=\sevenbb
\scriptscriptfont\bbfam=\fivebb
\def\bb{\fam\bbfam}

\def\Rb{{\bb R}}

\def\un{{\rm 1\mkern-4mu I}}

\def\build#1_#2^#3{\mathrel{
\mathop{\kern 0pt#1}\limits_{#2}^{#3}}}

\title{{\huge General Relativity and Experiment}\footnote{
To appear in the Proceedings of the XIth International Congress
of Mathematical Physics (Paris,July 1994)}}
\author{\Large Thibault Damour \\
\\
\normalsize Institut des Hautes Etudes Scientifiques \\
\normalsize 91440 Bures-sur-Yvette, France \\
\\
\small and \\
\\
\normalsize DARC, CNRS - Observatoire de Paris \\
\normalsize 92195 Meudon, France \\
\\}
\date{\small October 21, 1994}

\begin{document}

\maketitle

\bigskip

\begin{abstract}

The confrontation between Einstein's theory of gravitation and experiment is
summarized. Although all current experimental data are compatible with General
Relativity, the importance of pursuing the quest for possible deviations from
Einstein's theory is emphasized.

\end{abstract}

\newpage

\section{Introduction}

For many years, Einstein's General Relativity theory has been considered as a
mathematical structure rather than as a physical theory. This was due partly to
an insufficient recognition of the deep physical significance of the ``non
geometrical'' right-hand side of Einstein's equations [a ``shabby, wooden
construction'', as said Einstein, by contrast with the ``marble temple'',
geometrical left-hand side], and partly to the lack of experimental or
observational contacts of the theory. The situation has changed completely in
the last thirty years. From the conceptual point of view, it has been realized
through the work of many people [1] that the elegant, ``geometrical'' nature of
the left-hand side of Einstein's equations (containing the Ricci tensor)
followed, as a necessary consequence, from the physical postulate that the
source of gravity be the (inelegant, non geometrical) energy-momentum tensor.
 From the experimental and observational point of view, starting in the
sixties,
the implementation of high-precision (laboratory or spatial) tests of
Einstein's theory, and the discovery of new astrophysical objects (quasars, the
cosmic microwave background, pulsars,$\ldots$) obliged one to tackle in detail
the deep physical implications of General Relativity. In this brief review, we
summarize the current status of the confrontation between General Re\-lativity
and experiment with special emphasis on recent results, and we end by some
speculations about potentially fruitful improved experiments. For more details
and references we refer the reader to [2] or [3].

\bigskip

\section{Experimental tests of the coupling of matter to an external
gravitational field}

General Relativity can be thought of as defined by two postulates. One
postulate (equivalent to the choice of the geometrical left-hand side of
Einstein's field equations : $R_{\mu \nu} - \frac{1}{2} \ R \ g_{\mu \nu}$)
states that the action functional describing the propagation and
self-interaction of the gravitational field is proportional to the spacetime
integral of the curvature scalar $R$ of a pseudo-Riemannian [signature $-+++$]
four-dimensional manifold $(V_4 ,g_{\mu \nu})$:
$$
S_{\rm gravitation} \ [g_{\mu \nu}]= \frac{c^4}{16\pi \ G} \int \frac{d^4 x}{c}
\
\sqrt{g} \ R(g). \eqno (1)
$$
Here $g \equiv -\det \ g_{\mu \nu}$, and we use local coordinates $x^{\mu}
=(x^0 ,x^1 ,x^2 ,x^3)$. The second postulate (equivalent to the choice of the
non-geometrical right-hand side of Einstein's equations : $T_{\mu \nu}$, see
below) states that the action functional describing the coupling of all the
(fermionic and bosonic) fields describing matter and its electro-weak and
strong interactions is a (minimal) deformation of the special relativistic
action functional used by particle physicists (the so called ``Standard
Model''), obtained by replacing everywhere the flat Minkowski metric $f_{\mu
\nu} = {\rm diag} (-1,+1,+1,+1)$ by $g_{\mu \nu} (x^{\lambda})$ and the partial
derivatives $\partial_{\mu} \equiv \partial / \partial x^{\mu}$ by
$g$-covariant derivatives $\nabla_{\mu}$. [With the usual subtlety that one
must also introduce a field of orthonormal frames, a ``vierbein'', for writing
down the fermionic terms]. Schematically, one has
$$
S_{\rm matter} \ [\psi ,A,H,g] = \int \frac{d^4 x}{c} \ \sqrt{g} \ \Lc_{\rm
matter}, \eqno (2{\rm a})
$$
$$
\Lc_{\rm matter} = -\frac{1}{4} \sum \frac{1}{g_*^2} \ {\rm tr} (F_{\mu
\nu} \ F^{\mu \nu} ) - \sum \overline{\mathstrut \psi} \ \gamma^{\mu} \ D_{\mu}
\
\psi
$$
$$
\qquad \quad-\frac{1}{2} \ \vert D_{\mu} \ H \vert^2 - V(H) -\sum y \
\overline{\mathstrut \psi} \ H \ \psi , \eqno (2{\rm b})
$$
where $F_{\mu \nu}$ denotes the curvature of a $U(1)$, $SU(2)$ or $SU(3)$
Yang-Mills connection $A_{\mu}$, $F^{\mu \nu} =g^{\mu \alpha} \ g^{\nu \beta} \
F_{\alpha \beta}$, $g_*$ being a (bare) gauge coupling constant; $D_{\mu}
\equiv
\nabla_{\mu} +A_{\mu}$; $\psi$ denotes a fermion field (lepton or quark, coming
in various flavours and three generations); $\gamma^{\mu}$ denotes four Dirac
matrices such that $\gamma^{\mu} \ \gamma^{\nu} + \gamma^{\nu} \ \gamma^{\mu} =
2 g^{\mu \nu} \ \un_4$, and $H$ denotes the Higgs doublet of scalar fields,
with $y$ some (bare Yukawa) coupling constants.

\medskip

Einstein's theory of gravitation is then defined by extremizing the total
action functional,
$$
S_{\rm tot} \ [g,\psi ,A,H] = S_{\rm gravitation} \ [g] + S_{\rm matter} \
[\psi
,A,H,g] . \eqno (3)
$$
In particular, extremizing (3) with respect to $g_{\mu \nu}$ yields Einstein's
field equations
$$
R_{\mu \nu} - \frac{1}{2} \ R \ g_{\mu \nu} = \frac{8\pi \ G}{c^4} \ T_{\mu
\nu}
, \eqno (4)
$$
where $R_{\mu \nu}$ denotes the Ricci tensor, and $T_{\mu \nu} = g_{\mu \alpha}
\ g_{\nu \beta} \ T^{\alpha \beta}$, with
$$
T^{\mu \nu} \equiv \frac{2c}{\sqrt g} \ \frac{\delta \ S_{\rm matter}}{\delta \
g_{\mu \nu}} , \eqno (5)
$$
denotes the energy-momentum tensor of the ``matter''.

\bigskip

The second postulate of General Relativity, and more precisely the fact that
the matter Lagrangian (2b) depends only on $g_{\mu \nu} (x)$ and its first
derivatives, is a strong assumption which has many observable consequences for
the behaviour of (small) localized test systems embedded in given, external
gravitational fields. Indeed, from a theorem of Fermi and Cartan [4] stating
the existence of coordinate charts such that, along any given time-like curve,
the metric components can be set to their Minkowski values, and their first
derivatives made to vanish, follows the consequences:

\smallskip

\begin{enumerate}
\item[{\bf C}$_{\bf 1}$:] Constancy of the ``constants'' : the outcome of local
non-gravitational experiments depends only on the values of the coupling
constants and mass scales entering the laws of special relativistic physics.
[For instance, the cosmological time evolution has no influence on the value of
the fine-structure constant $\alpha = [137.0359895(61)]^{-1}$ which enters into
the physics of atoms].
\item[{\bf C}$_{\bf 2}$:] Local Lorentz invariance : local non-gravitational
experiments exhibit no preferred directions in spacetime. [In particular, the
local three-dimensional space is ``isotropic''].
\item[{\bf C}$_{\bf 3}$:] ``Principle of geodesics'' and universality of free
fall : small, electrically neutral, non self-gravitating bodies follow
geodesics of the external spacetime $(V,g)$. In particular, two test bodies
dropped at the same location and with the same velocity in an external
gravitational field fall in the same way, independently of their masses and
compositions.
\item[{\bf C}$_{\bf 4}$:] Universality of gravitational redshift : when
intercompared by means of electromagnetic signals, two identically constructed
clocks located at two different positions in a static external Newtonian
potential $U (\hbox{\bf x})$ exhibit, independently of their nature and
constitution, the difference in clock rate:
$$
\frac{\tau_1}{\tau_2} = \frac{\nu_2}{\nu_1} = 1 + \frac{1}{c^2} \ [U(\hbox{\bf
x}_1) - U(\hbox{\bf x}_2)] + O \left( \frac{1}{c^4}\right) . \eqno (6)
$$
\end{enumerate}

\medskip

Many experiments or observations have tested the observable consequen\-ces $C_1
-
C_4$ and found them to hold within the experimental errors. Many sorts of data
(from spectral lines in distant galaxies to a natural fission reactor
phenomenon which took place in Gabon two billion years ago) have been used to
set limits on a possible time variation of the basic coupling constants of the
Standard Model. The best results concern the fine-structure constant $\alpha$
for the variation of which a conservative upper bound is [5]
$$
\left\vert \frac{\dot{\alpha}}{\alpha} \right\vert < 10^{-15} \ {\rm yr}^{-1} ,
\eqno (7) $$
which is much smaller than the cosmological time scale $\sim 10^{-10} \
{\rm yr}^{-1}$.

\medskip

Any ``isotropy of space'' having a direct effect on the energy levels of atomic
nuclei has been constrained to the impressive $10^{-27}$ level [6]. The
universality of free fall has been verified at the $3\times 10^{-12}$ level for
laboratory bodies [7] and at the $10^{-12}$ level for the gravitational
accelerations of the Moon and the Earth toward the Sun [8]. The ``gravitational
redshift'' of clock rates given by eq. (6) has been verified at the $10^{-4}$
level by comparing a hydrogen-maser clock flying on a rocket up to an altitude
$\sim 10 \ \! 000$ km to a similar clock on the ground. Let us mention in
passing
that the general relativistic effect (6) is routinely taken into account in the
Global Positioning System which uses time signals from atomic clocks aboard
satellites to measure very accurately one's position on the Earth. This system
has been developed by the U.S. Army and finds now many practical civil \break
applications : accurate positioning of boats, airplanes, and, soon, of
indivi\-dual cars.

\bigskip

In conclusion, the main observable consequences of the Einsteinian postulate
concerning the coupling between matter and gravity have been verified with high
precision by all existing experiments. Therefore the simplest interpretation of
the present experimental situation is that the coupling between matter and
gravity is exactly of the form (2). We shall provisionally adopt this
conclusion to discuss the tests of the other Einsteinian postulate, eq. (1).
However, we shall come back at the end to the possibility of violations of eq.
(2).

\bigskip

\section{ Experimental tests of the dynamics of the gravitational field}

Let us now consider the experimental tests of the field equations (4) and, in
particular, of their left-hand side, i.e. tests of the dynamics of the
gravitational field defined by the action functional (1).

\bigskip

To discuss such tests it is convenient to enlarge our framework by
consi\-dering
the most natural relativistic theories of gravitation which would satisfy the
matter-coupling tests discussed in the previous section but differ in the
description of the degrees of freedom of the gravitational field. This class of
theories are the metrically-coupled (or mass-coupled) tensor-scalar theories in
which eq. (2) is preserved when written in terms of some ``physical'' metric
$g_{\mu \nu}$. The difference with General Relativity arises by demanding that
$g_{\mu \nu}$ be a composite object of the form
$$
g_{\mu \nu} = A^2 (\varphi) \ g_{\mu \nu}^* , \eqno (8)
$$
where the dynamics of the ``Einstein'' metric $g_{\mu \nu}^*$ is defined by the
action functional (1) (written with the replacement $g_{\mu \nu} \rightarrow
g_{\mu \nu}^*$) and where $\varphi$ is a massless scalar field. [More
generally, one can consider several massless scalar fields, with an action
functional of the form of a general nonlinear $\sigma$ model]. In other words,
the action functional describing the dynamics of the spin 2 and spin 0 degrees
of freedom contained in this generalized theory of gravitation reads
$$
S_{\rm gravitational} \ [g_{\mu \nu}^* ,\varphi ] = \frac{c^4}{16\pi \ G_*}
\int
\frac{d^4 x}{c} \ \sqrt{g_*} \ \left[R(g_*) - 2g_*^{\mu \nu} \ \partial_{\mu} \
\varphi \ \partial_{\nu} \ \varphi \right] . \eqno (9)
$$
Here, $G_*$ denotes some bare gravitational coupling constant. This class of
theories contains an arbitrary function, the ``coupling function''
$A(\varphi)$.
When $A(\varphi) = {\rm const.}$, the scalar field is not coupled to matter and
one falls back (with suitable boundary conditions) on Einstein's theory. The
simple, one-parameter subclass $A(\varphi) = \exp (\alpha_0 \ \varphi)$ with
$\alpha_0 \in \Rb$ is the Jordan-Fierz-Brans-Dicke theory. In the general
case, one can define the (field-dependent) coupling strength of $\varphi$ to
matter by $$\alpha (\varphi) \equiv \frac{\partial \ln A(\varphi)}{\partial
\varphi} . \eqno (10)
$$
It is possible to work out in detail the observable consequences of
tensor-scalar theories and to contrast them with the general relativistic case
(see ref. [10] for a recent treatment).

\bigskip

Let us first consider the experimental tests that can be performed in the solar
system. Because the planets move with slow velocities $(v/c \sim 10^{-4})$ in a
very weak gravitational potential $(U/c^2 \sim (v/c)^2 \sim 10^{-8})$, solar
system tests allow us only to probe the quasi-static, weak-field regime of
relativistic gravity (technically called the ``post-Newtonian'' limit). In this
limit, one does not explore the full structure of the gravitational theory but
only two of its ``Taylor coefficients'' in an expansion around the trivial flat
space solution. More precisely, one finds that all solar-system gravitational
experiments (with their current or foreseeable precision), interpreted within
tensor-scalar theories, differ from Einstein's predictions only through the
appearance of two ``post-Einstein'' parameters $\overline{\mathstrut \gamma}$
and
$\overline{\mathstrut \beta}$. These parameters vanish in General Relativity,
and are given in tensor-scalar theories by
$$
\overline{\mathstrut \gamma} = -2 \ \frac{\alpha_0^2}{1+\alpha_0^2} , \eqno
(11{\rm a}) $$
$$
\overline{\mathstrut \beta} = +\frac{1}{2} \ \frac{\beta_0 \
\alpha_0^2}{(1+\alpha_0^2)^2} , \eqno (11{\rm b})
$$
where $\alpha_0 \equiv \alpha (\varphi_0)$, $\beta_0 \equiv \partial \alpha
(\varphi_0) / \partial \varphi_0$; $\varphi_0$ denoting the
cosmologically-determined value of the scalar field far away from the solar
system. Essentially, the parameter $\overline{\mathstrut \gamma}$ depends only
on the linearized structure of the gravitational theory (and is a direct
measure
of its field content, i.e. whether it is pure spin 2 or contains an admixture
of
spin 0), while the parameter $\overline{\mathstrut \beta}$ parametrizes some of
the quadratic nonlinearities in the field equations (cubic vertex of the
gravitational field). All currently performed gravitational experiments in the
solar system, including perihelion advances of planetary orbits, the bending
and delay of electromagnetic signals passing near the Sun, and very accurate
range data to the Moon obtained by laser echoes, are compatible with the
general relativistic predictions $\overline{\mathstrut \gamma} = 0
=\overline{\mathstrut \beta}$ and give upper bounds on both $\left\vert
\overline{\mathstrut \gamma} \right\vert$ and $\left\vert \overline{\mathstrut
\beta} \right\vert$ (i.e. on possible fractional deviations from General
Relativity) of order $10^{-3}$ [8], [11].

\bigskip

In spite of the impressive quantitative value of solar system tests, one must
remember that they probe only the combined weak-field-quasi-stationary limit of
relativistic gravity. Fortunately, the discovery [12] and continuous
observational study of pulsars in gravitationally bound binary orbits has
provided nearly ideal laboratories for testing deeper aspects of relativistic
\break gravity : namely the propagation properties, and some of the
strong-field
structure, of the gravitational interaction. The reason why binary pulsars give
us a window on strong-field gravity is that they have a very strong
self-gravity, with surface potentials of order $GM/c^2 R \simeq 0.2$, i.e.
about a factor $10^8$ above the self-potential of the Earth, and a mere factor
$2.5$ below the black hole limit. The reason why they open a window on the
experimental study of the propagation properties of gravity, i.e. on its
radiative properties, is less evident. Heuristically, this is linked to an old
idea of Laplace [13] who argued that if gravity propagates with the velocity
$c_g$ the gravitational force acting on body $A$, member of a binary system,
should not be directed towards the instantaneous position of its companion $B$,
but should make a small angle $\theta \sim v/c_g$ away from it. This causes the
presence in the equations of motion of small terms of order $\theta$ times the
Newtonian $1/R^2$ force, directed against the velocities. These terms are
equivalent to damping forces; they cause the binary orbit to shrink and lead
therefore to a slow decrease in time of the orbital period $P_b : dP_b / dt
\sim - \theta$. The conclusion is that a careful monitoring of the orbital
period of a clean binary system (as is a binary pulsar) gives us access to the
lag angle $\theta$ due to the finite velocity of propagation of gravity. A
careful derivation of the equations of motion of binary systems of very compact
objects in General Relativity [14] has shown that the idea of Laplace was
morally correct, except for the fact that the lag angle $\theta$ is of order
$(v/c)^5$, which is numerically of order $10^{-12}$ in the case of the binary
pulsar ${\rm PSR} 1913+16$ ($c_g \equiv c$ in all relativistic theories of
gravity). More precisely, the general relativistic prediction for the orbital
period decay $\dot{P}_b \equiv dP_b / dt$ of a binary system of compact objects
of masses $m_1$ and $m_2$ is given by
$$
\dot{P}_b^{\rm GR} (m_1 ,m_2) = - \frac{192\pi}{5c^5} \ X_1 X_2 \ (GMn)^{5/3} \
\frac{P_4 (e)}{(1-e^2)^{7/2}} \eqno (12)
$$
where we have denoted
$$
M\equiv m_1 +m_2 \ , \ X_1 \equiv m_1 /M \ , \ X_2 \equiv m_2 /M \equiv 1-X_1 ,
$$
$$
n\equiv 2\pi / P_b \ , \ P_4 (e) \equiv 1 + \frac{73}{24} \ e^2 +
\frac{37}{96} \ e^4 .
$$
Due to the specific nonlinear structure of General Relativity the strong
self-gravitational effects of the compact objects (neutron stars or black
holes) do not appear explicitly in the formula (12) because they get
renormalized away in the definition of the (Schwarzschild) masses $m_1$ and
$m_2$.

\medskip

One sees from (12) that a measurement of $\dot{P}_b$ is not enough to provide a
test of General Relativity because one does not know beforehand the va\-lues of
$m_1$ and $m_2$. What is needed is the simultaneous measurement of at least
three ``post-Keplerian'' parameters (beyond the ``Keplerian'' ones such as the
binary period $P_b$ and the orbital eccentricity $e$ which appear also in (12))
which depend also on $m_1$ and $m_2$. The present observational situation
concerning the binary pulsar ${\rm PSR} 1913 + 16$ is that, thanks to the very
careful continuous experimental work of Taylor and collaborators over twenty
years, it has been possible to measure with accuracy three post-Keplerian
parameters, $\dot{P}_b$, $\dot{\omega}$ (periastron advance) and $\gamma$ (a
time dilation parameter, not to be confused with the post-Newtonian parameter
$\overline{\mathstrut \gamma}$). For instance, $\dot{P}_b^{\rm obs} =
-2.4225(56)
\times 10^{-12}$ [15] is known with the fractional precision $2.3 \times
10^{-3}$, which is an impressive achievement for such a small effect. The other
parameters, $\dot{\omega}^{\rm obs}$ and $\gamma^{\rm obs}$ are known with even
more precision. Each of these three parameters is predicted by General
Relativity to be a certain function of the two unknown masses $m_1$ and $m_2$.
In graphical terms, the simultaneous measurement of the three post-Keplerian
parameters $\dot{P}_b^{\rm obs}$, $\dot{\omega}^{\rm obs}$, $\gamma^{\rm obs}$
defines, when interpreted within the framework of General Relativity, three
curves in the $m_1 ,m_2$ plane, defined by the equations
$$
\dot{P}_b^{\rm GR} (m_1 ,m_2) = \dot{P}_b^{\rm obs} , \eqno (13{\rm a})
$$
$$
\dot{\omega}^{\rm GR} (m_1 ,m_2) = \dot{\omega}^{\rm obs} , \eqno (13{\rm b})
$$
$$
\gamma^{\rm GR} (m_1 ,m_2) = \gamma^{\rm obs} . \eqno (13{\rm c})
$$
These equations (where the explicit formulas for the functions
$\dot{\omega}^{\rm GR}$ and $\gamma^{\rm GR}$ will be found in, e.g., Ref. [3])
yield {\it one test of General Relativity}, according to whether the three
curves
meet at one point, as they should. As is discussed in detail in Refs. [15],
[16], [17], [18], General Relativity passes this test with complete success at
the $3.5 \times 10^{-3}$ level. [The final error being increased with respect
to the experimental error on $\dot{P}_b^{\rm obs}$ because of the necessity to
take into account a small perturbing effect caused by the Galaxy].

\bigskip

The success of the $\dot{P}_b - \dot{\omega} -\gamma$ test in ${\rm PSR} 1913 +
16$ is an impressive confirmation of General Relativity in a regime which has
not
been explored by solar system tests. The only reservation one can have about it
is that it represents an embarrassment of riches in that it probes, at the same
time, the radiative {\it and} the strong-field aspects of relativistic gravity
! Fortunately, the recently discovered binary pulsar ${\rm PSR} 1534 + 12$ [19]
has opened a new testing ground, in which it has been possible to probe
strong-field gravity independently of radiative effects.

\medskip

By fitting the observational data of ${\rm PSR} 1534 + 12$ to a generic
relativistic ``timing formula'' [20], it has been possible to measure four
independent post-Keplerian parameters, $\dot{\omega}$, $\gamma$, $r$ and $s$.
Each of these four parameters is predicted by General Relativity to be a
certain
function of the a priori unknown masses, $m_1$ and $m_2$, of the pulsar ${\rm
PSR} 1534+12$ and its companion. In graphical terms, the four simultaneous
measurements define four curves in the $m_1 - m_2$ mass plane of ${\rm PSR}
1534+12$. As these parameters involve strong-self-gravity effects but no
radiative effects, they provide $4-2=2$ tests of the strong-field regime of
relativistic gravity. As is discussed in detail in Ref. [17] (see also [15]),
General Relativity passes these {\it two} strong-field tests with complete
success.

\bigskip

To end this brief summary, let us mention that it has been possible to extend
the parametrization of eventual deviations from General Relativity (within the
general class of tensor-multi-scalar theories) by means of strong-field
parameters $\beta' , \beta'' , \beta_2 ,\ldots$ going beyond the weak-field
parameters $\overline{\mathstrut \gamma} , \overline{\mathstrut \beta}$
discussed above [10]. The comparison between binary pulsar data and the
predictions of some generalized gravitation theories has been made and has led
to significant bounds on the values of $\beta'$ and $\beta''$ [17]. Finally, a
comprehensive analysis of the maximum number of tests of gravitation theories
which can be extracted from binary pulsar data has been made [21], and has
concluded that, in principle (i.e. in the optimum experimental and
astrophysical
conditions) each binary pulsar system can provide {\it fifteen} tests of
relativistic gravity. Let us note that ${\rm PSR} 1534+12$ has recently
provided
a third test [22] (involving $\dot{P}_b$ and thereby giving an independent
confirmation of the reality of gravitational radiation) and might soon offer
the
possibility of seeing the relativistic spin precession induced by the
gravitational spin-orbit coupling [21].

\bigskip

\section{ Was Einstein 100\% right ?}

Summarizing the experimental evidence discussed above, we can say that
Einstein's postulate of a pure metric coupling between matter and gravity
appears to be, at least, $99.999 \ \! 999 \ \! 999 \ \! 9\%$ right (because of
universality-of-free-fall experiments), while Einstein's postulate (1) for the
field content and dynamics of the gravitational field appears to be, at least,
$99.9\%$ correct both in the quasi-static-weak-field limit appropriate to
solar-system experiments, and in the radiative-strong-field regime explored by
binary pulsar experiments. Should one apply Ockham's razor and decide that
Einstein must have been $100\%$ right, and then stop testing General Relativity
? My answer is definitely, no !

\bigskip

First, one should continue testing a basic physical theory such as Ge\-neral
Relativity to the utmost precision available simply because it is one of the
essential pillars of the framework of physics. Second, some very crucial
qualitative features of General Relativity have not yet been verified : in
particular the existence of black holes, and the direct detection on Earth of
gravitational waves. [Hopefully, the LIGO/VIRGO network of interferometric
detectors will observe gravitational waves early in the next century]. Last,
there are theoretical arguments suggesting that the interaction between
(electrically neutral) macroscopic bodies at low-energy might not be entirely
given by Einstein's theory. In other words, our current list of fundamental
interactions might not be complete, and there might exist some extra bosonic
field, with macroscopic range, mediating small but non zero forces between two
bodies. One such possibility is the existence of extra $U(1)$ vector fields
[23]. For instance, a field of range one meter, coupled to baryon number with
strength $\build <_{\sim}^{} 10^{-3}$ that of gravity would be compatible
with all the existing experimental evidence. A second possibility, is that the
cosmological evolution of the universe at large could have dynamically driven a
non-general-relativistic theory to a state where its predictions are very close
to the general relativistic ones [24]. In particular, it has been recently
suggested [25] that some of the gauge-neutral scalar fields appearing in string
theory as partners of $g_{\mu \nu}$ (the dilaton or a moduli) might exist in
the low-energy world today as very weakly coupled massless fields. As,
generically, such fields violate the metric-coupling postulate (2), this
provides a new motivation for trying to improve by several orders of magnitude
the various experimental tests of the observable consequences $C_1 - C_4$
discussed in section 2 above. In particular, this adds interest to the project
of a Satellite Test of the Equivalence Principle (nicknamed STEP, and currently
studied by ESA, NASA and CNES) which aims at probing the universality of free
fall of pairs of test masses orbiting the Earth at the (impressive) $10^{-17}$
level [26].

\newpage

\section*{References}

\begin{enumerate}
\item[{[1]}] R.P. Feynman, {\it Lectures on Gravitation}, unpublished lecture
notes (1962-1963) prepared by
F.B. Morinigo and W.G. Wagner (California Institute of Technology, 1971); \\
S. Weinberg, Phys. Rev. {\bf 138} (1965) B988, \\
V.I. Ogievetsky and I.V. Polubarinov, Ann. Phys. N.Y. {\bf 35} (1965) 167; \\
W. Wyss, Helv. Phys. Acta {\bf 38} (1965) 469; \\
S. Deser, Gen. Rel. Grav. {\bf 1} (1970) 9; \\
D.G. Boulware and S. Deser, Ann. Phys. N.Y. {\bf 89} (1975) 193; \\
J. Fang and C. Fronsdal, J. Math. Phys. {\bf 20} (1979) 2264; \\
R.M. Wald, Phys. Rev. D {\bf 33} (1986) 3613; \\
C. Cutler and R.M. Wald, Class. Quantum Grav. {\bf 4} (1987) 1267; \\
R.M. Wald, Class. Quantum Grav. {\bf 4} (1987) 1279.

\item[{[2]}] C.M. Will, {\it Theory and Experiment in Gravitational Physics},
2nd edition (Cambridge University Press, Cambridge, 1992); and Int. J. Mod.
Phys. D {\bf 1} (1992) 13.

\item[{[3]}] T. Damour, in {\it Gravitation and Quantizations}, eds B. Julia
and J. Zinn-Justin, Les Houches, Session LVII (Elsevier, Amsterdam, 1994).

\item[{[4]}] E. Fermi, Atti Accad. Naz. Lincei Cl. Sci. Fis. Mat. \& Nat. {\bf
31} (1922) 184 and 306; \\
E. Cartan, {\it Le\c cons sur la G\'eom\'etrie des Espaces de Riemann}
(Gauthier-Villars, Paris, 1963).

\item[{[5]}] P. Sisterna and H. Vucetich, Phys. Rev. D {\bf 41} (1990) 1034.

\item[{[6]}] J.D. Prestage et al., Phys. Rev. Lett. {\bf 54} (1985) 2387; \\
S.K. Lamoreaux et al., Phys. Rev. Lett. {\bf 57} (1986) 3125; \\
T.E. Chupp et al., Phys. Rev. Lett. {\bf 63} (1989) 1541.

\item[{[7]}] Y. Su et al., Phys. Rev. D {\bf 50} (1994) 3614.

\item[{[8]}] J.O. Dickey et al., Science {\bf 265} (1994) 482.

\item[{[9]}] R.F.C. Vessot and M.W. Levine, Gen. Rel. Grav. {\bf 10} (1978)
181; \\
R.F.C. Vessot et al., Phys. Rev. Lett. {\bf 45} (1980) 2081.

\item[{[10]}] T. Damour and G. Esposito-Far\`ese, Class. Quant. Grav. {\bf 9}
(1992) 2093.

\item[{[11]}] R.D. Reasenberg et al., Astrophys. J. {\bf 234} (1979) L219.

\item[{[12]}] R.A. Hulse and J.H. Taylor, Astrophys. J. Lett. {\bf 195} (1975)
L51.

\item[{[13]}] P.S. Laplace, {\it Trait\'e de M\'ecanique C\'eleste}, (Courcier,
Paris, 1798-1825), Second part : book 10, chapter 7.

\item[{[14]}] T. Damour and N. Deruelle, Phys. Lett. A {\bf 87} (1981) 81; \\
T. Damour, C.R. Acad. Sci. Paris {\bf 294} (1982) 1335; \\
T. Damour, in {\it Gravitational Radiation}, eds N. Deruelle and T. Piran
(North-Holland, Amsterdam, 1983) pp 59-144.

\item[{[15]}] J.H. Taylor, Class. Quant. Grav. {\bf 10} (1993) S167
(Supplement 1993).

\item[{[16]}] T. Damour and J.H. Taylor, Astrophys. J. {\bf 366} (1991) 501.

\item[{[17]}] J.H. Taylor, A. Wolszczan, T. Damour and J.M. Weisberg, Nature
{\bf 355} (1992) 132.

\item[{[18]}] J.H. Taylor, Phil. Trans. R. Soc. London A {\bf 341} (1992) 117.

\item[{[19]}] A. Wolszczan, Nature {\bf 350} (1991) 688.

\item[{[20]}] T. Damour and N. Deruelle, Ann. Inst. H. Poincar\'e {\bf 43}
(1985) 107 and {\bf 44} (1986) 263.

\item[{[21]}] T. Damour and J.H. Taylor, Phys. Rev. D. {\bf 45} (1992) 1840.

\item[{[22]}] A. Wolszczan and J.H. Taylor, to be published.

\item[{[23]}] P. Fayet, Phys. Lett. B {\bf 69} (1977) 489; B {\bf 171} (1986)
261; B {\bf 172} (1986) 363.

\item[{[24]}] T. Damour and K. Nordtvedt, Phys. Rev. Lett. {\bf 70} (1993)
2217; Phys. Rev. D {\bf 48} (1993) 3436.

\item[{[25]}] T. Damour and A.M. Polyakov, Nucl. Phys. B {\bf 423} (1994) 532;
Gen. Rel. Grav. {\bf 26} (1994), in press.

\item[{[26]}] P.W. Worden, in {\it Near Zero : New Frontiers of Physics}, eds
J.D. Fairbank et al. (Freeman, San Francisco, 1988) p. 766; \\
J.P. Blaser et al., {\it STEP Assessment Study Report}, ESA document SCI (94)5,
May 1994; \\
R. Bonneville, GEOSTEP, CNES document, DP/SC, April 1994.
\end{enumerate}

\end{document}